\documentclass[USenglish,oneside,twocolumn,natbib]{article}
\usepackage[utf8]{inputenc}
\usepackage[big]{dgruyter_NEW}
\usepackage{longtable}
\usepackage{booktabs} 
\usepackage{url}
\usepackage{comment}
\usepackage{color}
\usepackage{graphicx}
\graphicspath{{./images/}}
\usepackage[dvipsnames]{xcolor}
\usepackage[final]{changes}

\usepackage{placeins}
\usepackage{authblk}
\usepackage{etoolbox}
\makeatletter
\patchcmd\@combinedblfloats{\box\@outputbox}{\unvbox\@outputbox}{}{%
   \errmessage{\noexpand\@combinedblfloats could not be patched}%
}%
 \makeatother

\cclogo{\includegraphics{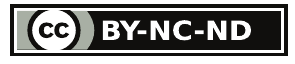}}
  
\begin{document}
 
  \author*[1]{Yan Shvartzshnaider$\dagger$}

  \author[2]{Noah Apthorpe$\dagger$}

  \author[3]{Nick Feamster}

  \author[4]{Helen Nissenbaum}

  \affil[1]{New York University \& Princeton University, E-mail: ys63@nyu.edu}
  
  \affil[2]{Princeton University, E-mail: apthorpe@cs.princeton.edu}

  \affil[3]{Princeton University, E-mail: feamster@cs.princeton.edu}

  \affil[4]{Cornell Tech, E-mail: helen.nissenbaum@cornell.edu\newline\textsuperscript{$\dagger$} These authors contributed equally to this work.}

\title{\huge Analyzing Privacy Policies Using Contextual Integrity Annotations}  

\runningtitle{Analyzing Privacy Policies Using Contextual Integrity Annotations}

\begin{abstract}
{In this paper, we demonstrate the effectiveness of using the theory of contextual integrity (CI) 
to annotate and evaluate 
privacy policy statements.
We perform a case study using CI annotations to compare Facebook's privacy policy before and after the Cambridge Analytica scandal. 
The updated Facebook privacy policy provides additional details about what information is being transferred, from whom, by whom, to whom, and under what conditions. 
However, 
some privacy statements prescribe an incomprehensibly large number of information flows by including many CI parameters in single statements. 
Other statements result in incomplete information flows due to the use of vague terms or omitting contextual parameters altogether.
We then
demonstrate that crowdsourcing can effectively produce CI annotations of privacy policies at scale.
We test the CI annotation task on 48 excerpts of privacy policies from 17 companies with 141 crowdworkers. The resulting high precision annotations
indicate that crowdsourcing could be used to produce a large corpus of annotated privacy policies for future research.
}
\end{abstract}

  \journalname{}
  \startpage{1}

  \journalyear{..}
  \journalvolume{..}
  \journalissue{..}
 
\maketitle


\section{Introduction}

Many online services operate by collecting and sharing users' information. To protect consumers, the U.S.~Federal Trade Commission (FTC) devised fair information practice principles (FIPPs) based on the  ``notice and choice'' framework~\cite{federal1998privacy}. These principles, in concert with state regulations, require companies to notify consumers about their information collection and sharing practices through privacy policies. These privacy policies, which often include details about the type of information collected, the entities that receive or store the information, and the conditions governing data acquisition and handling, serve two main purposes: 1) informing consumers about data collection practices, which they can consider when deciding whether or not to use a service, and 2) offering regulators, such as the FTC, a way to audit online services for misleading privacy practices. 

As we write this paper, the European General Data Protection Regulation (GDPR)~\cite{euGDPR} is coming into effect, forcing companies to adapt their behavior and rewrite their privacy policies or face strict penalties. The changes are largely based on GDPR Articles 13, 14, and 15, which outline the details companies need to provide to consumers when collecting, processing and sharing their information. The regulation puts an emphasis on providing this information to the ``subject in a concise, transparent, intelligible and easily accessible form, using clear and plain language''~\cite{Article12GDPR}. As a result, consumers are receiving an avalanche of updated privacy policies as companies strive for GDPR compliance~\cite{GDPREmails}.  However, just because the GDPR has pushed companies to update their privacy policies does not necessarily mean that these updated policies address the issues of previous versions. 

In this paper, we make a case for using the theory of contextual integrity (CI)~\cite{nissenbaum2010privacy}  to annotate, assess, and compare information sharing practices disclosed in privacy policies, both within and across updates. We showcase this technique with a case study in which we use the CI framework to manually annotate Facebook's previous and updated privacy policies to  identify the {\em senders, recipients} and {\em subjects} of information, information types {\em (attributes)}, and the conditions under which information may be transferred or collected ({\em transmission principles}). We then use these annotations to gain insight into the privacy policy and amendments. 

Our analysis shows that while the updated privacy policy includes statements that describe almost as twice as many information flows as the current policy, it fails to provide more clarity to the consumer. In many cases, the updated policy has more incomplete and ambiguous information flow statements than the current policy. 
Incomplete information flow statements (45\% of all statements in the current policy and 63\% in the updated policy) do not mention one or many information flow parameters. 
This allows readers to interpret the missing parameters according to their own expectations, which may not match the actual practices of the company. 
In contrast, some statements in both current and updated policies suffer from what we refer to as ``parameter bloating,'' i.e., they contain more than one instance of each CI parameter. 
This increases the cognitive load required for consumers to fully comprehend all possible information flows allowed by the statement. 
Finally, we identified privacy statements (over 50\% in both current and updated policies) that use vague and ambiguous language.

To help streamline our approach beyond the Facebook case study, we present a methodology for crowdsourcing CI privacy policy annotations. We construct CI annotation as a Amazon Mechanical Turk (AMT) Human Intelligence Task (HIT) and compare crowdsourced annotations against ground-truth expert annotations. 
We test the annotation task on 48 excerpts of privacy policies from 17 companies with 141 AMT workers. The crowdsourced annotations have an average word-based precision score of $0.9$ across CI information flow parameters.
This high precision indicates that CI annotation is both understandable and easily applicable by those with no prior exposure to CI, despite the often legalistic language employed by privacy policies.
This provides further evidence that CI successfully expresses how most people intuitively reason about information privacy. 
Finally, the high precision of crowdsourced annotations indicates that crowdsourcing could be applied at scale to evaluate future privacy policy updates or to build a dataset for training a machine learning model to perform automatic CI annotations. 

In summary, this work makes the following contributions:
\begin{enumerate}
    \item We present a method for annotating privacy policies using the contextual integrity framework. The use of a structured framework allows rigorous analysis of difficult privacy policy texts that is applicable to policies across companies and sectors.
    \item We describe a case study using CI annotation to analyze recent updates to Facebook's privacy policy, which identifies several issues with information flow descriptions across versions. 
    \item We demonstrate that crowdsourcing can produce precise CI annotations of legalistic privacy policy excerpts for future CI annotation research at scale.
\end{enumerate}

\section{CI Primer}

The theory of CI is based on two central premises: 1) privacy is defined as the appropriateness of information flows, which 2) is defined by contextual norms governing  particular settings (contexts) in which information is transmitted \cite{nissenbaum2010privacy}. 

CI offers a template for describing information flows using 5-parameter tuples, which include specific actors ({\em senders}, {\em recipients}, and {\em subjects}) involved in the information flow, the type ({\em attribute}) of the information, and the condition ({\em transmission principle}) under which the information flow occurs.  
This combination of five parameters defines contexts which determine privacy norms. For example, while someone might consider sharing  Fitbit\footnote{https://www.fitbit.com/home} data with their doctor, they might view the sharing of this same data with advertising or insurance companies 
as a privacy violation. The entire context, including recipient and information type, affects how we think about privacy. 

The CI framework was previously used as a lens for examining android permissions~\cite{wijesekera2015android}, online platform  practices~\cite{hull2011contextual,zimmer2008privacy}, and examining GDPR regulations~\cite{guinchard2017contextual} themselves. In more recent efforts, CI was employed to capture individuals' privacy expectations, which can be then checked for inconsistencies or used to inform policymakers and manufacturers~\cite{apthorpe2018discovering,shvartzshnaider2016learning}.
\section{Related work}

Privacy policies are notoriously hard to read. As a result, average users find them difficult to comprehend and correctly interpret. This leads to gaps between users' expectations and the stated policy~\cite{martin2015privacy}. 

The problem of privacy policy comprehension has been the focus of many previous studies. Some efforts focused on lexical~\cite{sathyendra2016automatic, evans2017evaluation} and semantic~\cite{sathyendra2017identifying} analysis of the privacy policies. Others works~\cite{Wilson2016} used crowdsourcing to provide annotations that allow users to more easily parse privacy policies and identify sections related to specific concerns, as well as to help researchers assess policies from different websites.

The Usable Privacy Policy Project (UPPP)~\cite{sadeh2013usable} has recruited law students to hand-annotate 115 privacy policies with metadata tags such as ``first party collection/use,'' ``user choice/control,'' ``data retention,'' and ``data security.'' They then used the hand-labeled policies to train a machine learning algorithm that has annotated over 7,000 policies with the same metadata tags~\cite{wilson2016creation}. 
While extracting relevant paragraphs saves time for the interested reader, it does not provide a way of identifying issues with the policy itself. It is remains up to the reader to interpret the text. This tends to create gaps between privacy expectations and policy statements, especially when policy statements are ambiguous or incomplete~\cite{martin2015privacy}.

Recent work has shown evidence that privacy policies often elide or obscure crucial contextual information that could help users formulate their privacy expectations. 
In 2016, Martin and Nissenbaum~\cite{martin2016measuring} showed that when confronted with a privacy-related scenario that was missing some contextual information, respondents mentally supplemented the information, essentially generating a different version of the scenario.  Martin and Nissenbaum also conducted a survey of 569 respondents presented with 40 scenarios with random combination of contextual factors. The results showed that the ``context of information exchange -- how information is used and transmitted, the sender and receiver of the information -- all impact the privacy expectations of individuals''~\cite{martin2016measuring}. 

The importance of including contextual factors was also reported by Rao et al., in a 2016 study that compared users' privacy expectations with existing companies' practices~\cite{RaoUsenixExptn2016}.  
240 participants were asked to state their expectations for the data collection, sharing, and deletion practices of 16 websites across finance, health, and dictionary categories. The results showed that users' privacy expectations depend on the type of website and the type of information being exchanged. For example, respondents expected medical data to be shared with a medical website, but not a financial website. These findings provide further evidence to support the importance of contextual factors in how individuals perceive privacy practices, motivating a contextual analysis of privacy policies to identify gaps which might result in mismatched privacy expectations.

Another body of work has explored using crowdsourcing to annotate privacy policies, thereby splitting the cognitive load of understanding an individual policy over multiple workers. 
In 2016, Wilson et. al.,~\cite{Wilson2016} explored the feasibility of  asking crowdworkers to answer questions on data collection practices. In the experiment, 218 crowdworkers were assigned the task of reading through 12 privacy policies and answering 9 questions about data collection, sharing, and deletion practices stated in the policies. To support their answers, respondents needed to annotate the relevant text in the privacy policies. The results showed that the answers of the crowdworkers agreed with those of skilled annotators over 80\% of the time. The results indicate that crowdsourcing can be used to identify paragraphs describing specific practices in privacy policies. 
Our results support this conclusion, but extend it to even more sophisticated annotations of individual components of contextual information flows described in privacy policies.

\section{Annotation Methodology} \label{sec:annotation-methodology}

We use the CI framework to annotate policy statements that describe contextual information exchanges.
Our use of a CI flow-based abstraction is an important distinction from previous privacy policy annotation research, as it serves a useful semantic abstraction for checking privacy statements for more complex properties than previously attempted.
For the remainder of the paper, we denote a privacy statement with a single set of CI parameters as an ``information flow.'' For example, we consider the following statement an information flow, or simply as a ``flow:''
 \medskip\\
 \noindent\fbox{%
    \parbox{0.45\textwidth}{%
    {\em  {\bf \textcolor{ForestGreen}{We [Facebook]}}$^{\textcolor{red}{recipient}}$ also collect {\bf \textcolor{blue}{contact information}}$^{\textcolor{red}{attribute}}$ that {\bf \textcolor{orange}{you}}{$^{\textcolor{red}{sender}}$} provide {\bf \textcolor{Fuchsia}{if you upload, sync or import this information (such as an address book) from a device.}}$^{\textcolor{red}{TP}}$} }%

}
\medskip\\ 
This flow contains an explicit sender, recipient, attribute, and transmission principle. The subject parameter is not included, but is implicitly the consumer agreeing to the privacy policy. 

We use the following guidelines to identify CI parameters within individual flows for annotation:
\begin{itemize}
\item {\bf Sender.} Any entity (person, company, website, device, etc.) that transfers or shares the information. This may be a pronoun or a specific entity, such as ``Company A,'' ``strategic partners,'' or ``publisher.''

\item {\bf Recipient.} Any entity (person, company, website, device, etc.) that ultimately receives the information. This may be a pronoun or a specific entity, such as ``third party,'' ``developer,'' ``other users,'' or ``Company B and its affiliates.''

\item {\bf Transmission principle.} Any clause describing the ``terms and conditions under which [...] transfers ought (or ought not) to occur''~\cite{nissenbaum2010privacy}. This includes descriptions of how information may be used or collected. Examples include ``if the user gives consent,'' ``when an update occurs,'' or ``to perform specified functions.''

\item {\bf Attribute.}  Any description of information type, instance, and/or example, such as ``date of birth,'' ``credit card number,'' ``photos,'' or, more generally, ``personal information.''

\item {\bf Subject.} Any subjects of the information exchanged in a flow. Subjects may be explicitly stated or implicitly described using pronouns and possessives.

\end{itemize}

\begin{figure}[t]
\centering
\includegraphics[width=0.45\textwidth]{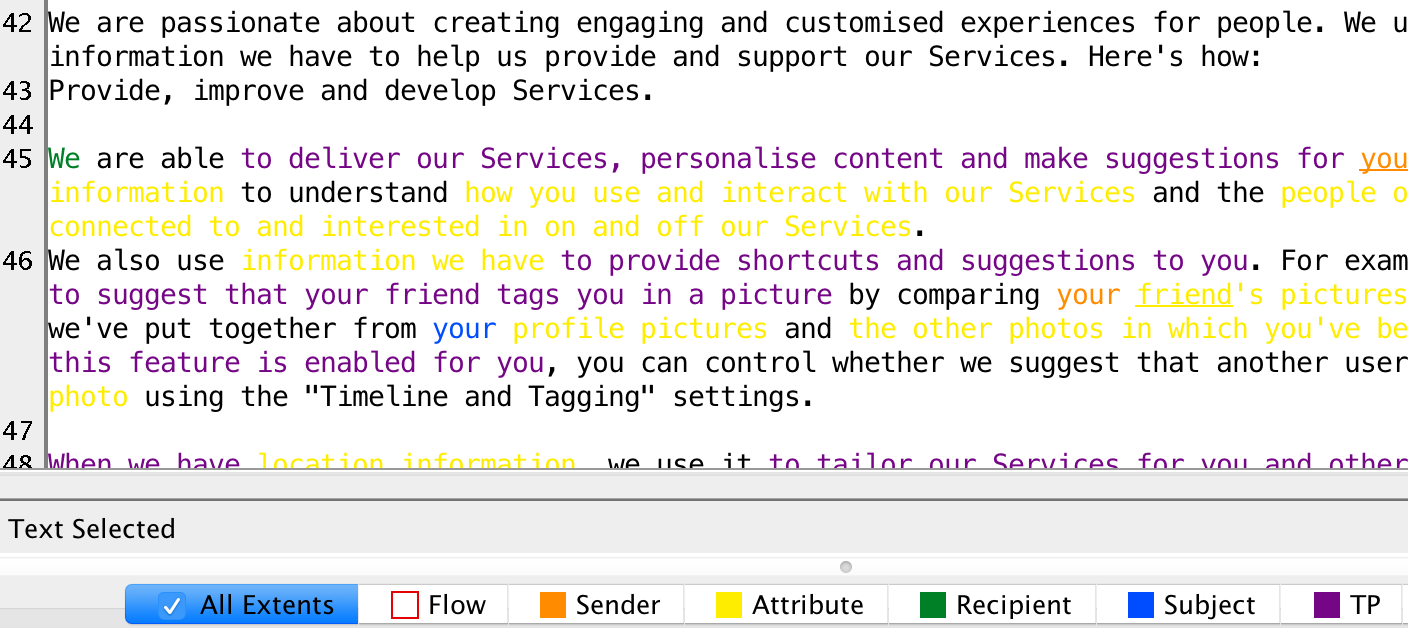}
\caption{Screenshot of the Multi-document Annotation Environment tool configured for CI privacy policy annotation.  }
\label{fig:annotation_tool}
\end{figure}

We perform manual policy annotation using Document Type Definitions (DTD) markup and the Multi-document Annotation Environment~\cite{MAT} (Figure~\ref{fig:annotation_tool}).

\section{Facebook Case Study}\label{sec:case_study}
Recent revelations about the misuse of consumer data by Facebook and Cambridge Analytica~\cite{bloomberg_story} has rekindled the debate around users' privacy and informed consent on such platforms. Facebook claims~\cite{wpost} that they  provide users with the right level of control to keep their information private. They also claim that consumers are well informed by the disclosure of information handling practices in the company's privacy framework. Assuming that this is indeed the case, i.e., ignoring the complexity and a sporadic evolution of Facebook controls~\cite{lipford2008understanding,liu2011analyzing,johnson2012facebook}, we see the Cambridge Analytica scandal as another example of how things can go wrong when consumers' privacy expectations are misaligned with privacy policy statements.

Much of the problem stems from not having a coherent higher-level abstraction that can help reason about privacy policies. While talented legal scholars and professionals are trained to identify relevant privacy policy excerpts and mentally stitch them into coherent flows, so to speak, the average consumer is usually overwhelmed by the legal language of privacy policies~\cite{turow2008consumers}. Even experts themselves find some privacy policy statements confusing~\cite{reidenberg2015disagreeable}. 

Furthermore, research shows that consumers' tend to  ``[project] the important factors to their privacy expectations onto the privacy notice''~\cite{martin2015privacy}. In other words, consumers implicitly fill in the blanks left by difficult-to-interpret policies, which inadvertently widens the gap between their expectations and actual company behaviors.

As a result of public outcry~\cite{bloomberg_story}, Facebook has amended its privacy policy to include a more detailed account of its information sharing practices.
It is therefore timely and instructive to apply our CI annotation technique to the previous and updated Facebook privacy policies in order to demonstrate the power of the method and highlight issues with both policy versions. 

\subsection{Analysis}

Using the methodology described in Section~\ref{sec:annotation-methodology}, we manually annotated Facebook's previous privacy policy (data policy)
as well as the official updated version\footnote{https://www.facebook.com/about/privacy/update}.
The following sections demonstrate the range of analyses that can be performed using CI annotations but are not exhaustive. We anticipate a variety of additional analytic techniques building on these annotations in future work. 

\begin{table*}
\centering
\begin{tabular}{p{1.8cm}p{4cm}p{9.5cm}}
  CI Parameter &  Previous & Updated \\ 
    \midrule
   Sender 
   & people you share and communicate with &  specific friends or accounts, friends and followers, other people using Facebook and Instagram, people \\\\ 
   & devices, phones, computers,  devices where you install or access our Services  & connected TVs, web-connected devices you use that integrate with our Products\\
     \midrule
    Recipient & family of companies that are part of Facebook & Facebook companies, Facebook company products \\\\
    & people you share and communicate & audience they choose, specific friends or accounts, those you connect and share with around the world, people in your networks,  friends and followers, people and businesses outside the audience that you shared with, anyone who can see the other person's content, anyone on or off our products\\\\
    & partners conducting academic research, partners conducting  surveys &  research partners, research partners who we collaborate with, academics \\\\
    & third-party companies who help us provide and improve our services or who use advertising or related products & websites that integrate with our products, other services that integrate with our products, companies that aggregate	\\\\
    & N/A & systems, devices and operating systems providing native versions of Facebook and Instagram (i.e. where we have not developed our own first-party apps), anyone on or off our product, content creator, seller, page admins, regulators, network  \\
    \midrule
    Attribute & information about how you use our services, how you use and interact with our services & information about any of your Instagram followers, the ads you see and how you use their services, other web-connected devices you use that integrate with our products, when you last used our products, whether a window is foregrounded or backgrounded,  when you're using and have last used our products, identifiers from apps or accounts that you use, actions that you have taken on our products\\\\
    & content about you & the features you use, life events, racial or ethnic origin, activities, where you live, what games you play, information about your interests actions and connections, who you are ``interested in", your health, events you attend, interests, preferences, your religious views, general demographic, the places you like to go and the businesses and people you're near, whether you are currently active on Instagram messenger or Facebook, check-ins, websites you visit, other information about your Facebook friends from you, political views, trade union membership, philosophical beliefs\\\\
    & information about the reach and effectiveness of their advertising & reports about the kinds of people seeing their ads, which Facebook ads led you to make a purchase or take an action with an advertiser, ads you see, family device ids\\\\
    & Device information & information about operations and behaviours performed on the device, other identifiers unique to Facebook company products associated with the same device or account, available storage space\\\\
    & N/A  & information about nearby wi-fi access points beacons and cell towers \\
    \midrule
    Transmission Principle & N/A & to detect when someone needs help, to recognise you in photos videos and camera experiences, help you stream a video from your phone to your tv, combat harmful conduct, can help distinguish humans from bots, to aid relief efforts, whether or not you have a Facebook account or are logged in to Facebook, reshared or downloaded through APIs, to have lawful rights to collect, use and share your data before providing any data to us and many others. 
\\
 \end{tabular}
\caption{List of notable CI parameters that were introduced or refined between the previous and updated Facebook privacy policies.}
\label{tbl:compared_params}
\end{table*}

\subsubsection{Comparison of CI parameters}

We compared the number information flows prescribed by both previous and updated Facebook privacy policies (Figure~\ref{fig:FB_summary}) and the CI parameters they contain. We matched CI parameters across policies using fuzzy string matching~\cite{cohen2011fuzzywuzzy} with the following thresholds for each CI parameter: sender (70\%), attribute (65\%), recipient (70\%), and transmission principle (55\%). While the fuzzy string matching worked well, some corner cases required manual validation. We describe some notable differences between information flows in the previous and updated policies on a parameter-by-parameter basis as follows:

\begin{figure}[t]
\centering
\includegraphics[width=0.45\textwidth]{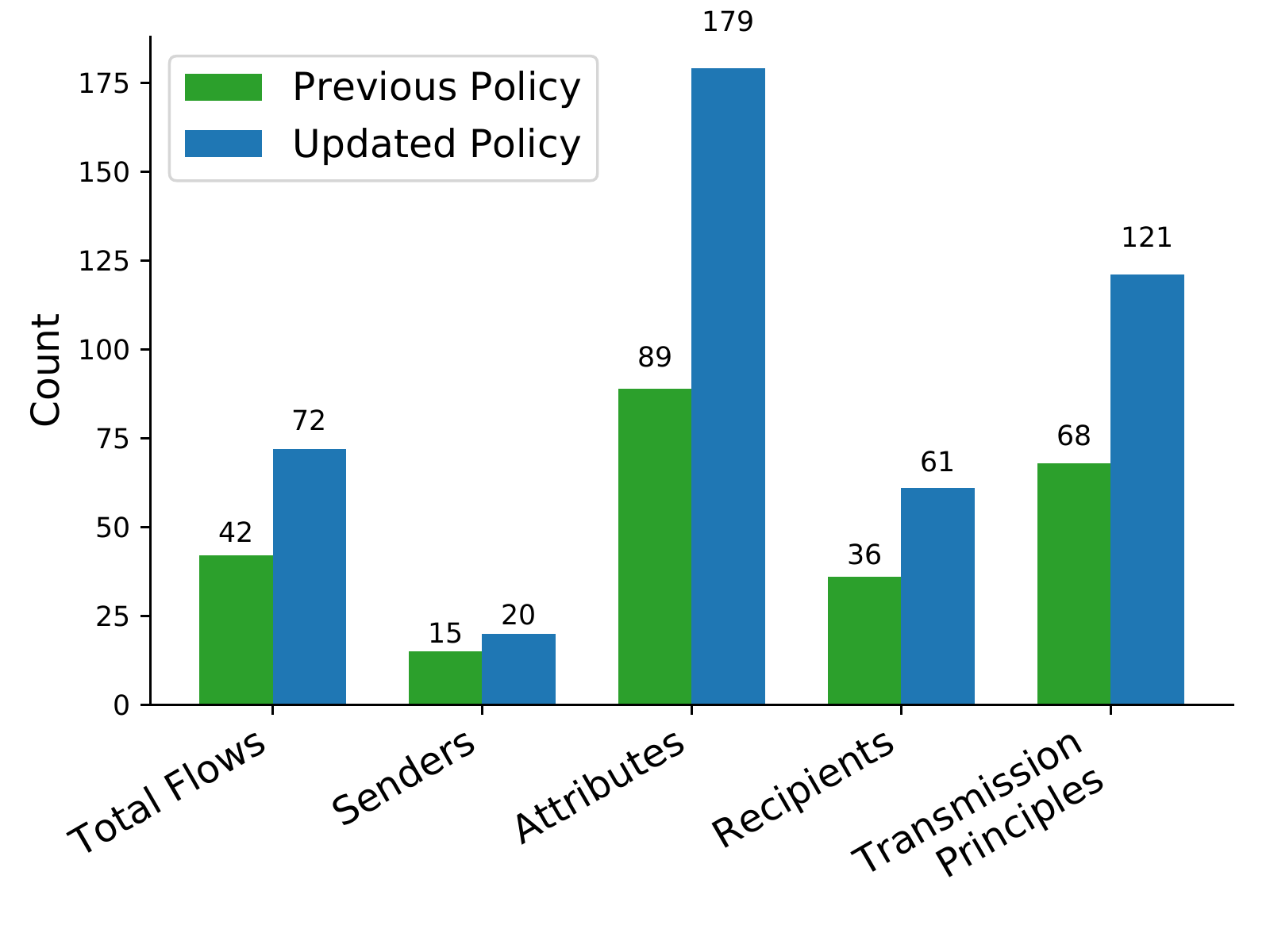}
\vspace{-1.8em}
\caption{Distribution of unique CI parameters identified in the previous and updated Facebook privacy policies.}
\label{fig:FB_summary}
\end{figure}

\textbf{Sender.} The updated policy offers a more detailed account of the sources of information transfer. It elaborates on categories from the previous privacy policy and also includes several new senders, such as ``WhatsApp'', ``connected TV'', ``a business'' which were not specified in the previous policy.
Not surprisingly, the most frequent senders in both policies are Facebook and the consumer (Table~\ref{tbl:param_freq_summary}).

\textbf{Recipient.} Similarly to the sender parameter,  the updated version introduces new recipients, such as ``people and businesses outside the audience that you shared with,'' ``content creators,'' ``page admin,'' ``Instagram business profiles,'' and ``companies that aggregate.''  
As expected, the most common ``recipients'' in both versions are ``Facebook,'' and  ``third party service, vendors, partners'' (Table~\ref{tbl:param_freq_summary}).

\begin{table}[t]
\centering
\begin{tabular}{llp{4.5cm}}
   CI Param & Version & Instances (frequency)  \\ 
\midrule
Recipients & Previous & we [Facebook]  (22), Third party service, vendors, partners (20) \\
& Updated & we [Facebook]  (32), Third party service, vendors, partners (24) \\
&&\\
Senders & Previous  &  we [Facebook]  (14), you (6) \\
& Updated & we [Facebook] (17),  you (11)  \\
&&\\
Attributes & Previous  &  information (8), information about you (2), information we have (2), non-personally identifiable information only (2)  \\
& Updated &  information (15), content (5), information about you (4), information that we have (4), public information (4), communications (2), shipping and contact details (2). \\
\end{tabular}
\caption{The most frequent recipients, senders, and attributes mentioned in the previous and updated Facebook privacy policies.}
\label{tbl:param_freq_summary}
\end{table}

\textbf{Attribute.} When describing the types of information being transferred or collected, the updated policy contains more attributes (179) than the the previous policy (86). However, we note that some attributes from the previous policy were omitted in the update. The updated policy does not mention ``user id'' (opting for ``username'' instead), or ``age range'' (instead providing the example ``\dots ad was seen by a woman between the ages of 25 and 34''). 
Generally, the updated policy describes new types of information and/or elaborates on information that was previously generic or abstract (Table~\ref{tbl:compared_params}). For example, the updated draft provides significantly more details about the type of content that is being collected about the user, including ``racial or ethnic origins,''  ``health,'' ``events attended,'' ``interests,'' ``religious views,'' ``general demographics,'' ``political views,'' ``trade union membership,'' and ``philosophical beliefs.'' Furthermore, the updated policy describes attributes not discussed in the previous policy, such as ``connected TVs,'' ``information about nearby Wi-Fi access points,'' ``beacons,'' and ``cell towers.''

\textbf{Transmission Principle.} When specifying conditions under which  information transfer may be performed, the updated policy includes all conditions and information flow constraints in the previous policy. In addition, the updated policy also contains new transmission principles, such as  ``whether or not you have a Facebook account or are logged in to Facebook,'' ``to recognise you in photos, videos and camera experiences,'' ``reshared or downloaded through APIs,'' ``to have lawful rights to collect, use and share your data before providing any data to us,'' and many others (Table~\ref{tbl:compared_params}). 

\textbf{Subject.}
The subject of most flows in both policies is the consumer. We therefore do not include the subject parameter in our analysis. 

\subsubsection{Incomplete Information Flows} \label{sec:missing}
\begin{figure}[t]
\centering
\includegraphics[width=0.45\textwidth]{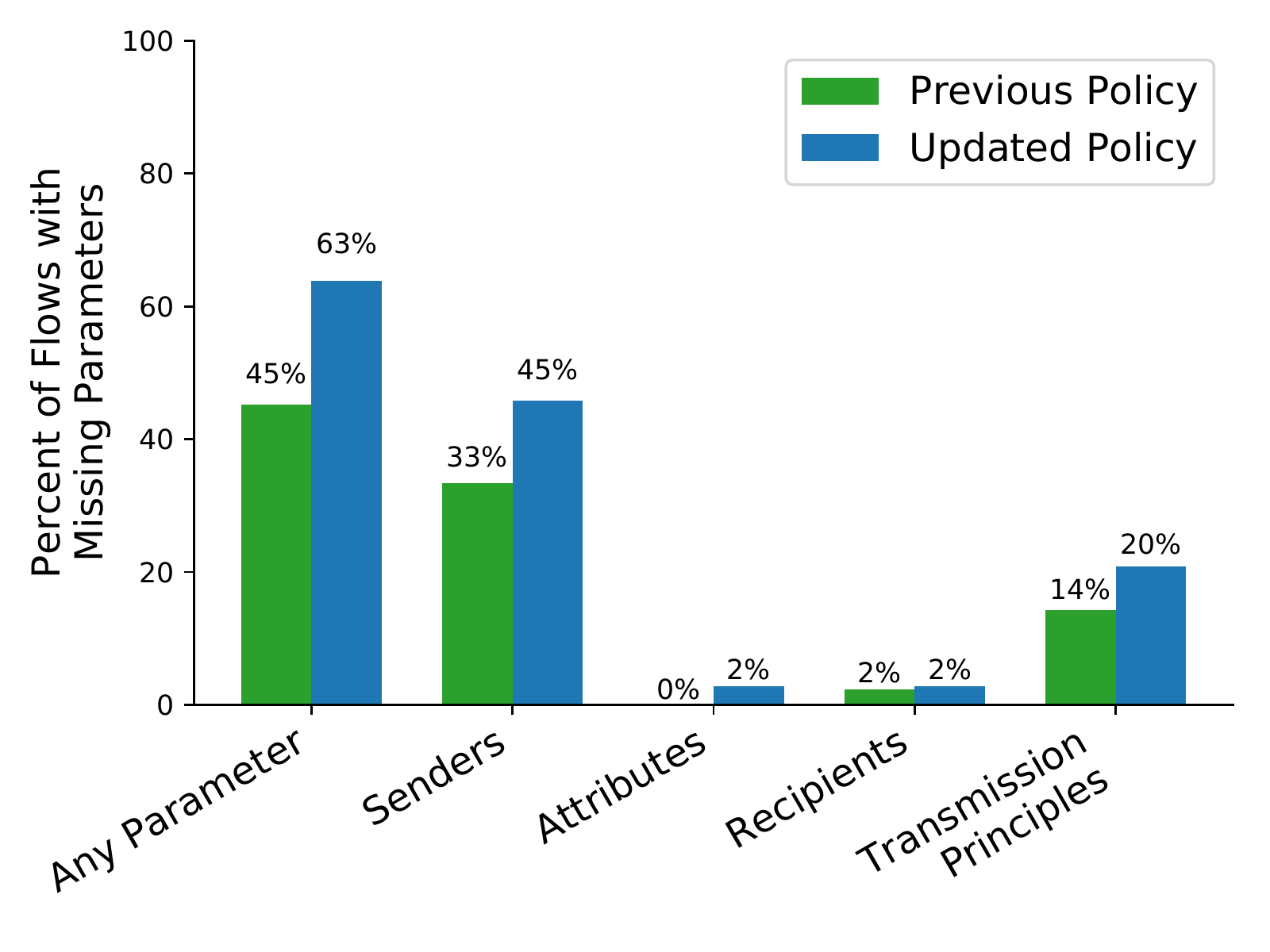}
\vspace{-1.4em}
\caption{Percentage of incomplete information flows in Facebook's previous and updated privacy policies with missing CI parameters.}
\label{fig:FB_missing_params}
\end{figure}

Our analysis of the Facebook privacy policies finds many prescribed information flows with missing (non-specified) parameters (Figure~\ref{fig:FB_missing_params}). 
Failing to specify parameters introduces ambiguity, leaving consumers uninformed about company behavior.
In the previous privacy policy, $45\%$ (19/42) of flows are missing one or more parameters. In the updated policy, this number increases to $68\%$ (49/72). 

{\bf Missing Recipient.} Table~\ref{tbl:missing_rcpt} lists  the flows from both policies with missing recipient parameter. The previous policy only has one flow without an explicit recipient while the updated policy has two. Not stating information recipients forces users to infer what entities will have access to their information from other sources, often leading to incorrect notions of company behavior~\cite{turow2018persistent, martin2016measuring}. Identifying the recipient can sometimes be difficult, as in the flow {\em``We are able to suggest that your friend tags you in a picture by comparing your friend's pictures to information we've put together from your profile pictures and the other photos in which you've been tagged.''}

\begin{table}[t]
\centering
\begin{tabular}{p{6cm}c}
   Information Flow & Version \\ 
   \midrule
   When you comment on another person's post or like their content on Facebook, that person decides the audience who can see your comment or like & Previous \\
   &\\
   You can choose to provide information in your Facebook profile fields or life events about your religious views, political views, who you are ``interested in'' or your health. This and other information (such as racial or ethnic origin, philosophical beliefs or trade union membership) could be subject to special protections under the laws of your country & Updated\\
   &\\
For example, people can share a photo of you in a story or mention, tag you at a location in a post or share information about you in their posts or messages & Updated \\
\end{tabular}
\caption{Information flows in the previous and updated Facebook privacy policies with missing recipient parameters.}
\label{tbl:missing_rcpt}
\end{table}

\textbf{Missing Sender.} The sender parameter is not specified in 14 ($33\%$)  flows in the previous policy nor in 33 ($45\%$) flows in the updated policy. Many of the statements with missing senders describe ``use-of-data,'' i.e., they inform the consumer how the collected information will be used but not from where it is collected. Missing senders can easily lead to misinterpretations and false privacy expectations. 
For example,  the source of the information in the following statement is unclear: {\em ``We collect information about the people, Pages, accounts, hashtags and groups you are connected to and how you interact with them across our Products, such as people you communicate with the most or groups you are part of.''} 

\textbf{Missing Transmission Principle.} We identified 6 information flows in the previous policy where the transmission principle was missing. For example, the statement ``{\em We share information we have about you within the family of companies that are part of Facebook}'' does not specify under what conditions/constraints the information is being shared. Likewise, the statement ``{\em We also collect information about how you use our Services, such as the types of content you view or engage with or the frequency and duration of your activities. Things others do and information they provide}'' does not contain any transmission principles. These statements force consumers to guess when and for what reason information is collected. 

The updated policy contains even more (15) flows with missing transmission principles. Without a transmission principle, flows like ``{\em We also receive information about your online and offline actions and purchases from third-party data providers who have the rights to provide us with your information''} become ambiguous because it is not clear when or why this information is being collected.

\subsubsection{CI Parameter Bloating}
Our CI annotation analysis also identifies several flows in both previous and updated policies with multiple CI parameters of the same type. We refer to this phenomenon as  CI parameter bloating.
Parameter bloating adds to the cognitive effort required to isolate single information flows from privacy policy statements,
because it is often not clear which combinations of parameters describe information flows that actually take place.

Consider the following flow from the updated policy:

\medskip
\noindent\fbox{%
    \parbox{0.45\textwidth}{%
    {\bf \textcolor{Orange}{Advertisers, app developers and publishers}}$^{\textcolor{red}{senders}}$ can send \textcolor{ForestGreen}{us}$^{\textcolor{red}{recipient}}$ information {\bf \textcolor{Fuchsia}{through Facebook Business Tools that they use, including our social plug-ins (such as the Like button), Facebook Login, our APIs and SDKs or the Facebook pixel}}{$^{\textcolor{red}{TP}}$}. These partners provide information about {\bf your}{$^{\textcolor{red}{subject}}$} {\bf \textcolor{blue}{activities off Facebook including information about your device, websites you visit, purchases you make, the ads you see and how you use their services whether or not you have a Facebook account or are logged in to Facebook}}$^{\textcolor{red}{attributes}}$.}%
}
\medskip

\noindent 
At first glance, the above privacy statement seems transparent and informative. It explicitly specifies the type of information that is being exchanged, between what actors and under what conditions. 
However, this is an example of CI parameter bloating. The prescribed information flow is overloaded with CI parameters. Note the many senders (advertisers, app developers and publishers) attributes (information about your device, websites you visit, purchases you make, the ads you see and how you use their services), and transmission principles (when you use Like, Facebook login, APIs, SDKs and through Facebook Pixel). How does the consumer reason about this information flow? Do all listed senders transfer all of these information types to Facebook or does each particular sender transmit a specific information type? Do  flows with each sender/information pair occur under each listed TP or only specific ones? Even technically-savvy users will have difficulty reasoning about the many possible information flows with all combinations of each parameter type.

We would like to emphasize that specifying multiple instances of the same parameter does not automatically lead to  parameter bloating.
Specifically, parameter bloating does not include instances where a single parameter is enumerated to clarify a given category, as in the following statement, which elaborates on several attributes:

\medskip
\noindent\fbox{%
    \parbox{0.45\textwidth}{%
    { {\em \textcolor{ForestGreen}{We}$^{\textcolor{red}{recipients}}$ collect \textcolor{blue}{information about how use our Products, such as types of content you view or engage with, the features you use, the actions you take, the people or accounts you interact with and the time, frequency and duration of your activities}}$^{\textcolor{red}{attributes}}$.}}
    }%

\begin{figure}[t]
\centering
\includegraphics[width=0.5\textwidth]{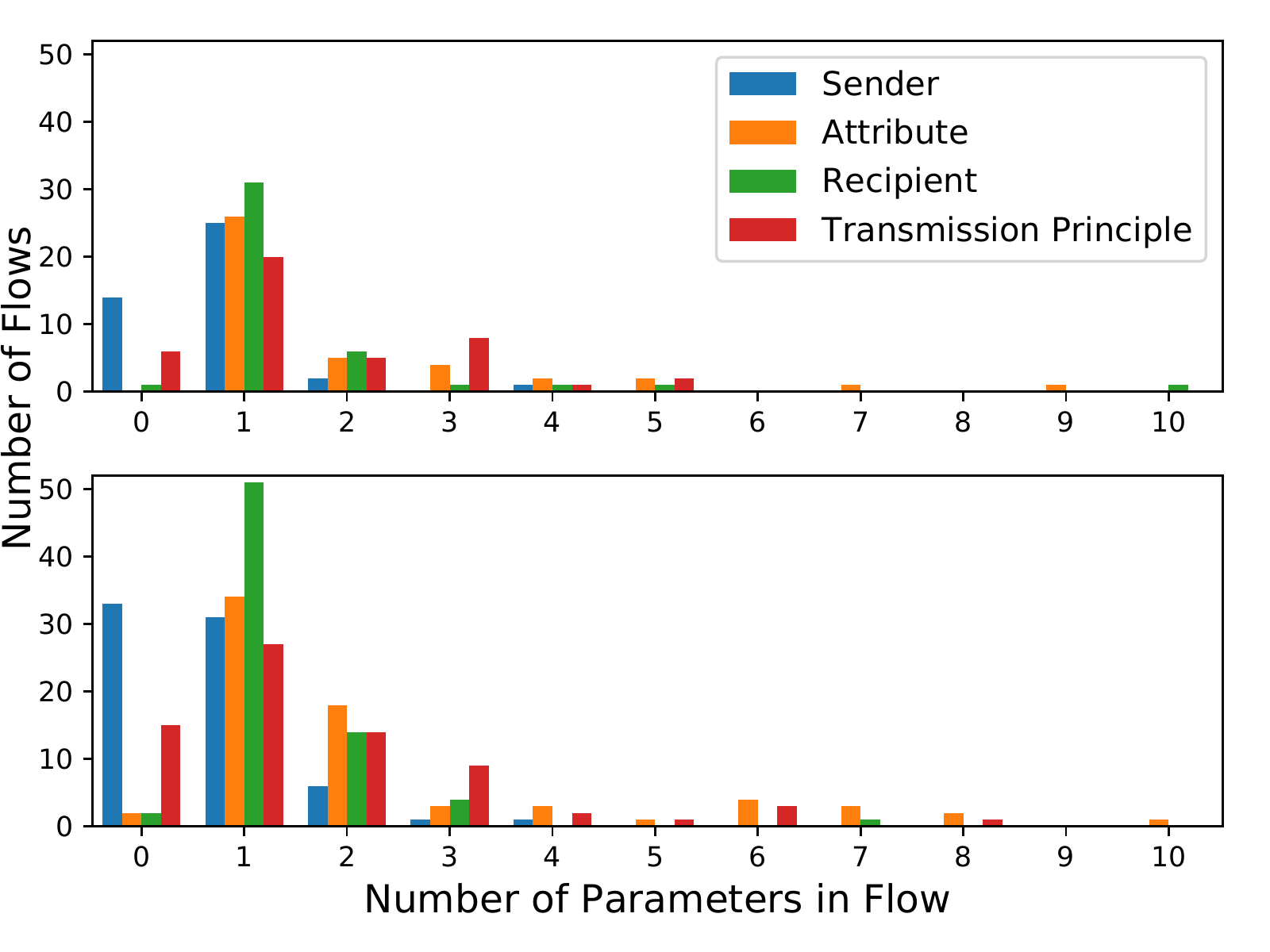}
\caption{Number of CI parameters per flow in Facebook's previous (\textit{top}) and updated (\textit{bottom}) privacy policies. The previous policy had one flow with 18 attributes and the updated policy has one flow with 40 attributes that are omitted for readability.}
\label{fig:param_frequency}
\end{figure}

\medskip
Figure~\ref{fig:param_frequency} shows the number of  CI parameters per flow in both current and updated policies. In the previous policy, there are 10 information flows that mention more than one recipient, with one information flow standing out, listing 10 potential recipients. Three flows mention more than one sender, and 16 flows mention multiple attributes, ranging  from 2 to 18 attributes per flow. Multiple transmission principles appear in 16 flows, ranging from 2 to 5 TPs per flow. 

The updated policy contains even more bloated flows.  Multiple senders appears in 8 information flows (2 senders in 6 flows, 3 in 1 flow, and 4 in 1 flow). Multiple attributes occur in 36 flows ranging from 2 attributes in 18 flows to 40 attributes in a single flow.  Nineteen of the flows include more than one recipient (2 recipients in 14 flows, 3 in 4 flows, and 7 in 1 flow). Finally, the number of flows with multiple transmission principles increased to 30, ranging from 2 TPs in 14 flows to 8 TPs in a single flow.

Given that an average consumer today spends little to no time reading privacy policies, it is unreasonable to assume that the even the most privacy-concern citizen will dissect all possible combinations of this many multi-parameter flows. 

\subsubsection{Vague and Ambiguous Flows}\label{sec:vague}

CI annotation analysis also allow us to identify information flows
that use vague terminology as defined in~\cite{bhatia2016theory} (Table~\ref{tbl:catgeories_of_vaguness}).  

\begin{table*}[t]
\centering
\begin{tabular}{lp{5cm}p{8cm}}
  {\bf Category} & {\bf Definition} &  {\bf Example Terms} \\ 
\midrule
Conditionality &  it is not clear what is the condition associated with information transfer &              
            ``as needed'', 
            ``as necessary'',
             ``as appropriate'',
             ``depending'',
             ``sometimes'',
             ``as applicable'',
             ``otherwise reasonably determined'',
             ``from time to time''\\
& &  \\                    
Generalization & action or information types are too abstract or vague & ``typically", 
                  ``normally", 
                  ``often" , 
                  ``general",
                  ``usually",
                  ``generally",
                  ``commonly ",
                  ``among other things", 
                  ``widely", 
                  ``primarily",
                  ``largely",
                  ``mostly" \\
& & \\               
Modality & 	Hard to estimate the possibility of occurrence & ``likely",
            ``may",
            ``can",
            ``could" 
            ``would",
            ``might",
            ``could",
            ``possibly" \\
& & \\                     
Numeric Quantifier & Vague numeric quantifier & ``certain", 
                      ``most",
                      "majority", 
                      "many", 
                      "some"
                      "few" \\
 \end{tabular}
 \vspace{8pt}
\caption{Summary of four vagueness categories as defined in~\cite{bhatia2016theory} and associated example terms.}
\label{tbl:catgeories_of_vaguness}
\end{table*}

\begin{figure}[t]
    \centering
    \includegraphics[width=0.45\textwidth]{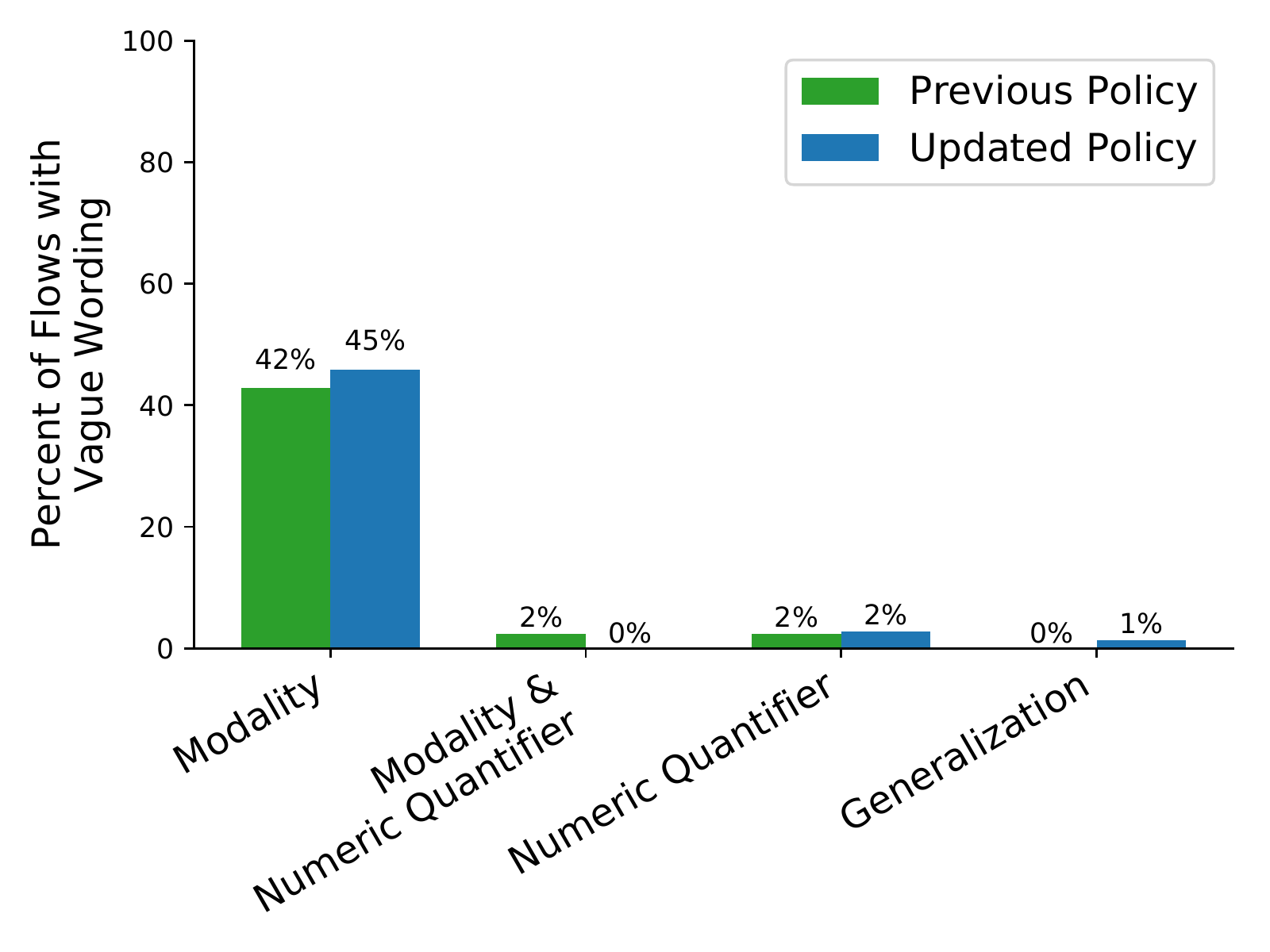}
    \vspace{-1.5em}
    \caption{Percentage of information flows in Facebook's previous and updated privacy policies with various categories of vague wording (categories defined in Table~\ref{tbl:catgeories_of_vaguness})}.
    \label{fig:FB_vagueness}
\end{figure}

Figure~\ref{fig:FB_vagueness} shows the percentage of flows in Facebook's previous and updated policies that use vague terminology. In both policies, ``modality'' vagueness dominates, occurring in close to 45\% of all flows. The updated policy does not represent a reduction in vague terminology from the previous version. 
Rather, the percentage of flows with vague terms remains the same. 
This supports our initial claim the updated data policy does not contribute to clarity. 
The widespread occurrence of flows with vague wording further supports the problem that privacy policies are too often ``obtuse and noncommittal [and] make it difficult for people to know what information a site collects  and how it will be used''~\cite{turow2008consumers}. 

\subsection{Summary}
The updated Facebook privacy policy has twice as many information flows as the previous policy (Figure~\ref{fig:FB_summary}). However, more information flows does not necessarily equal less confusion. Our analysis shows that many of the newly introduced information flows are either incomplete, overloaded with CI parameters and/or use vague terms.

Rather than fix fundamental issues in their privacy policy in the wake of the Cambrige Analytica scandal, Facebook seems to have opted to add more terms, entities, and conditions. While this may initially seem to provide additional information to the consumer, CI annotation analysis reveals that there are still many issues preventing users from interpreting clear information flows from these new details and from understanding how their data is being collected and shared. 

\section{Crowdsourcing CI Annotations}\label{sec:crowdsourcing}

The ability to effectively crowdsource CI annotation would allow researchers to efficiently pursue two primary goals: 1) collect a large dataset of annotations in order to train a machine learning model to perform CI annotation automatically, and 2) perform a large-scale analysis of information flows across the privacy policies of many companies. This would provide a broad sense of information flow disclosure practices across the technology sector via many of the same analysis methods used in the Facebook case study.

We have developed a crowdsourcing technique that poses CI annotation as an Amazon Mechanical Turk (AMT) Human Intelligence Task (HIT). We crowdsourced the annotation of 48 privacy policy excerpts, including 16 excerpts from the Google privacy policy circa October 2017 and 16 pairs of excerpts from the pre-GDPR and post-GDPR privacy policies of 16 well-known companies\footnote{Amazon, Fitbit, Indiegogo, LinkedIn, The New York Times, Mirosoft, Shapeways, Slack, Spotify, Steam, Stripe, Tinder, Twitter, Uber, WhatsApp, Yelp}. 
This choice of policy excerpts provides evidence that our crowdsourcing technique is effective within a single policy as well as for privacy policies across the technology sector. 
The excerpt pairs were selected as representative statements from the pre-GDPR policies of each company and the corresponding statements from the GDPR-compliant version of each policy updated in May 2018. The excerpts ranged from 21 to 113 words\footnote{Mean: 55 words/excerpt, SD: 23 words/excerpt} and from 1 to 4 sentences for a total of 2621 annotated words over 103 annotated sentences.

We compared the crowdsourced annotations to ground-truth annotations from a CI expert. The crowdsourced annotations had an average word-based precision of 0.9 across CI information flow parameters, indicating that the  crowdworkers understood the relatively complex notion of information flow parameters and were able to correctly identify them in real privacy policy text. 
These results show that crowdsourcing can be an effectual tool for CI annotation. 
We will release the crowdsourced annotations as a public dataset for further analysis upon publication.

Sections \ref{sec:task-design}--\ref{sec:crowdsource-discuss} describe the design and evaluation of our CI annotation crowdsourcing method in more detail. 

\subsection{Annotation Task Design}
\label{sec:task-design}

We developed the annotation task as a Qualtrics \cite{qualtrics} survey deployed on AMT. The task was designed to optimize annotation accuracy while minimizing cost. 

\textbf{Consent and Instructions.}
The first page of the annotation task is a consent form. Participants who do not consent are prevented from proceeding. The annotation task collects no personal information about crowdworkers and was approved by our university's Institutional Review Board. 

\begin{figure*}
\section*{Appendix}
    \centering
    \includegraphics[width=0.9\textwidth]{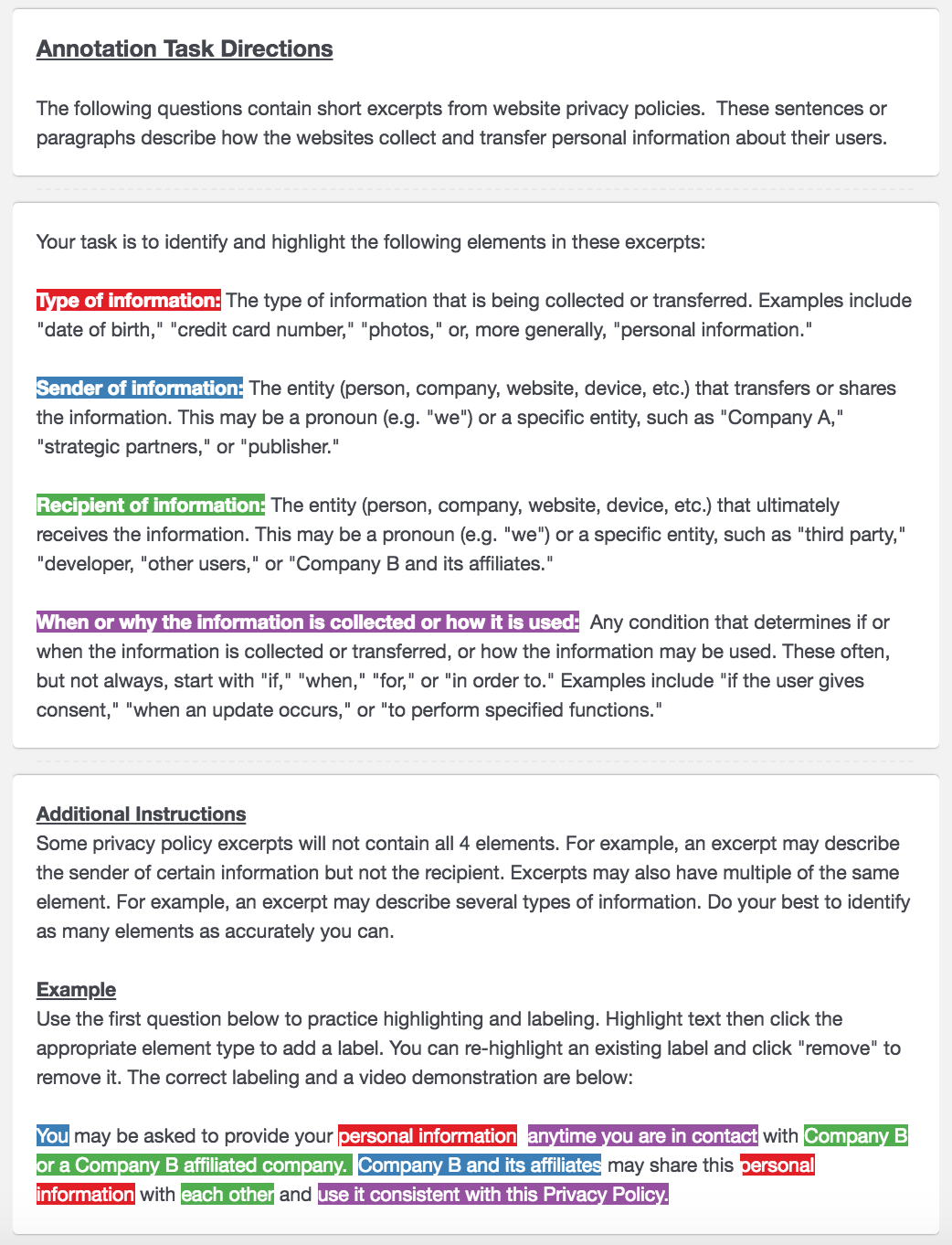}
    \caption{CI annotation task instructions.}
    \label{fig:task-instructions}
\end{figure*}

The task next presents annotation instructions  (Appendix Figure~\ref{fig:task-instructions}), including a description of each information flow parameter that should be annotated (sender, attribute, recipient, and transmission principle) and an example annotated flow. The information flow parameter descriptions match those used by expert annotaters as described in Section~\ref{sec:annotation-methodology}. 

\begin{figure}[t]
\centering
\includegraphics[width=0.48\textwidth]{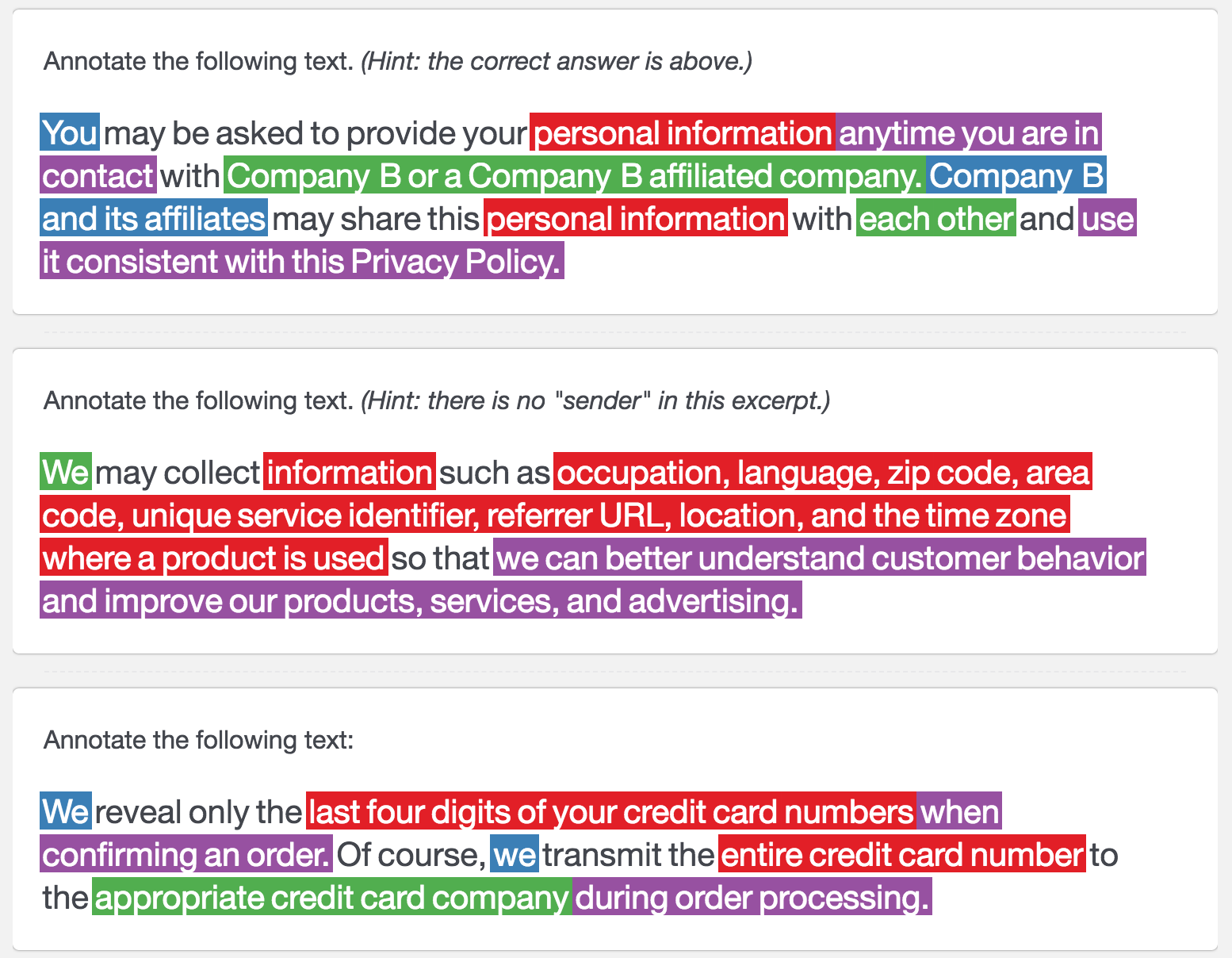}
\caption{Screening questions to identify AMT workers who are able to perform high accuracy annotations. The ground truth annotations are shown with sender in blue, recipient in green, attribute in red, and transmission principle in purple. Common English stopwords (except ``you,'' ``your,'' ``them,'' and ``we'') are not counted when comparing crowdworker annotations to the ground truth.}
\label{fig:screener-questions}
\end{figure}

\textbf{Screening Questions.}
Each crowdworker is asked to annotate (highlight and label) all words and phrases corresponding to CI information flow parameters in three privacy policy excerpts (Figure~\ref{fig:screener-questions}). These excerpts serve as screening questions to identify workers who are able to perform high-accuracy annotations. Workers whose annotations have at least a 0.7 word-based F\textsubscript{1} score (Section~\ref{sec:eval-metrics}) compared to ground-truth expert annotations on the first screening question (for which the correct answer is given) and either of the next two screening questions are allowed to proceed with the task. Workers whose annotations do not meet this accuracy threshold do not proceed. This helps limit the effect and cost of workers who do not understand the task or who attempt to ``cheat'' by performing minimal annotations (e.g., highlighting just the first word in each excerpt).  

\textbf{Annotations.} 
Each worker who passes the screening questions is then asked to annotate 5 of the 48 excerpts of interest, although these could be replaced with arbitrary privacy policy excerpts for future research.  
The format of these annotation questions is equivalent to the screening questions (Figure~\ref{fig:screener-questions}). The instructions are also repeated at the top of the page for workers to refer to if they wish. 

Annotations of all excerpts from multiple workers are collected, analyzed, and processed into the final crowdsourced annotation for each privacy policy (Section~\ref{sec:maj-vote}). 

The task concludes with a field for optional open-ended comments if participants have anything they wish to communicate to the researchers.

\subsection{Task Deployment}

We first tested the annotation task on UserBob~\cite{userbob}, a usability-testing service where users narrate their experience while performing tasks. We collected seven UserBob responses. All UserBob workers completed the task in less than 15 minutes. We used the UserBob responses to adjust task instructions to ameliorate worker confusion. Performing such ``cognitive interviews'' is common practice in survey design and development \cite{sudman1997thinking}.

We deployed the annotation task as a HIT on AMT using TurkPrime \cite{litman2017turkprime}, an online tool for researchers to easily manage AMT tasks. 
We limited the HIT to AMT workers in the United States with an HIT approval rating of 90--100\% and at least 100 HITs approved. 
141 total workers accepted the HIT. Of these workers, 99 passed the screener questions. All 48 excerpts were annotated by between 7 and 12 workers (mean~10.2). 
AMT workers who did not pass the screening questions were automatically reimbursed \$0.25. AMT workers who passed the screening test and completed the entire annotation task were reimbursed \$1.50. 
Collecting all responses took approximately 4 hours.

\subsection{Majority Vote Annotations}
\label{sec:maj-vote}

We are ultimately interested in acquiring the single highest-accuracy annotation for each privacy policy independent of individual workers. We therefore combine multiple annotations of each privacy policy excerpt into a ``majority vote'' annotation, which assigns each word in an excerpt to the CI parameter annotated by at least 50\% of the participants presented with that excerpt.
If fewer than 50\% of workers labeled a word with the same parameter, then the word is given no label in the majority vote annotation.

\subsection{Evaluation Metrics}
\label{sec:eval-metrics}
We had one of the authors perform expert ground truth annotations of all  excerpts prior to seeing the crowdsourced results.
We use the following evaluation metrics to compare the crowdsourced majority vote annotations to the expert annotations. 

\textbf{Parameter-based scoring.} 
We manually counted all instances of each CI parameter labeled in both the crowdsourced majority vote and expert annotations (true positives), in the expert annotation only (false negatives), and in the crowdsourced annotation only (false positives). 
We further categorized the false positives and false negatives  to better understand crowdworker mistakes and how to improve the annotation task in future studies (Section~\ref{sec:errors}).

\textbf{Word-based scoring.}
We also applied a automated word-based scoring method that did not require manually comparing variable-length parameters and could be used to easily evaluate future large-scale CI annotation efforts.

We first removed common English stopwords from all annotations to prevent variations in article or preposition highlighting from affecting annotation comparisons. We used the stopword list in Python NLTK library \cite{bird2009natural} less ``you,'' ``your,'' ``them,'' and ``we,'' as these pronouns could have been senders or recipients in the privacy policy excerpts.

True positives are then words labeled by both the participant and the expert. False positives are words labeled by the participant only. False negatives are words labeled by the expert only. 
This allows us to calculate word-based precision, recall, and F\textsubscript{1} scores for each CI parameter and excerpt.
Some CI parameters do not occur in every excerpt. 
If the expert did not label a particular parameter in an excerpt, participants' recalls were defined as 1 for the corresponding annotation. 
If a participant did not label a particular element in an excerpt, the participant's precision was defined as 1 for the corresponding annotation.
These are standard definitions of precision and recall for edge cases. 

\subsection{Annotation Accuracy}

\begin{figure}[t]
\centering
\includegraphics[width=0.48\textwidth]{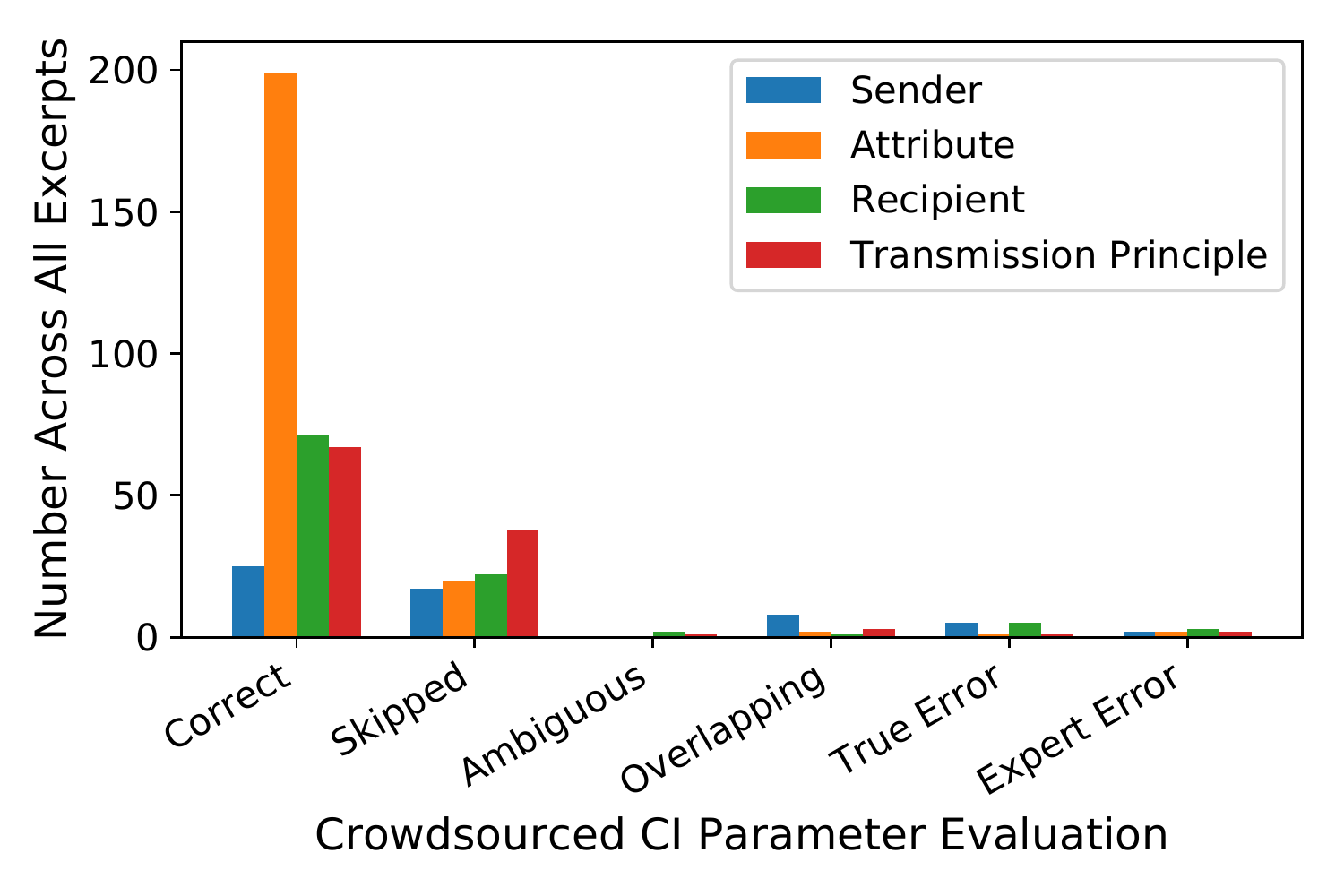}
\caption{Parameter-based evaluation of crowdsourced majority vote annotations compared to expert ground truth. Correct (true positive) annotations are parameters labeled to match the expert annotation. Skipped (false negative) annotations are parameters only labeled by the expert. All other incorrect annotations (false positives) are described in Section~\ref{sec:errors}. Note that most errors are skipped parameters (false negatives), indicating that the crowdworkers understood the task, but that further work is needed to improve recall. 
}
\label{fig:manual-scores}
\end{figure}

Figure~\ref{fig:manual-scores} shows the counts of correctly and incorrectly annotated CI parameters across all excerpts from parameter-based scoring. The incorrect annotations are divided into categories to better understand the source of crowdworker errors. 
The crowdsourced majority vote annotations correctly labeled 43\% of the senders, 89\% of the attributes, 68\% of the recipients, and 60\% of the transmission principles across all excerpts. False negatives were by far the most common error, with the crowdsourced annotations missing 30\% of the senders, 9\% of the attributes, 21\% of the recipients, and 34\% of the transmission principles across all excerpts. 
Finally, false positive errors comprised 26\% of the senders, 2\% of the attributes, 11\% of the recipients, and 6\% of the transmission principles across all excerpts.\footnote{Percentages were rounded to the nearest whole value and may not add to 100\%} 

\begin{figure}[t]
\centering
\includegraphics[width=0.48\textwidth]{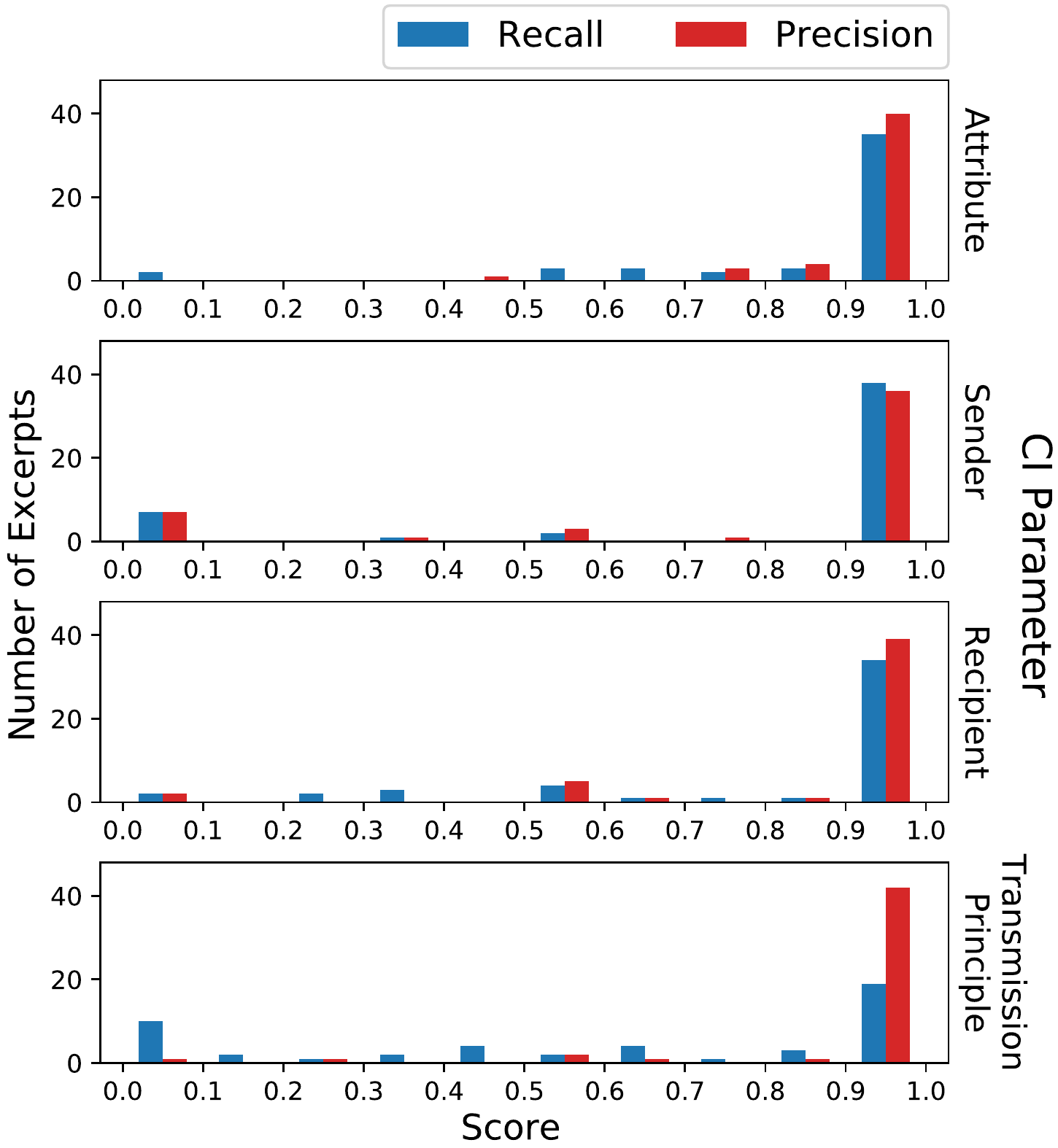}
\caption{Word-based precision and recall scores of majority vote crowdsourced annotations compared to expert ground truth for each CI element.}
\label{fig:turkers-maj-pr}
\end{figure}

Figure~\ref{fig:turkers-maj-pr} shows the distributions of word-based precision and recall scores for the majority vote annotations across all excerpts and for each CI parameter. The average precision across all excerpts is 0.95 for attributes, 0.80 for senders, 0.89 for recipients, and 0.94 for a transmission principles. The corresponding average recall across all excerpts is 0.87 for attributes, 0.82 for senders, 0.83 for recipients, and 0.59 for transmission principles. 

Overall, the high precision of the majority vote crowdworker annotations (by both parameter-based and word-based scoring methods) indicates that the majority of crowdworkers understood the CI annotation task, and were able to correctly identify and highlight CI parameters in short privacy policy excerpts. However, the many false negatives indicates that the framing of the task could potentially be improved to help crowdworkers avoid missing or intentionally skipping some parameters. 

\subsection{Evaluating Crowdworker Errors}
\label{sec:errors}

Analyzing the crowdsourced annotations raises the question ``What causes particular excerpts or CI parameters to be more difficult for crowdworkers to annotate than others?''

One intuitive explanation is that excerpts that are longer, more difficult to read, or contain more CI parameters are more difficult for crowdworkers to annotate. To test this hypothesis, we calculated Spearman correlations of the majority vote annotation word-based F\textsubscript{1} scores versus text length, Flesch-Kincaid Reading Ease \cite{kincaid1975derivation}, FOG Index \cite{kincaid1975derivation}, and number of CI parameters (Appendix Table~\ref{tbl:readability}).
However, all of the resulting correlation coefficients had absolute values less than $0.5$, indicating no strong correlations with F\textsubscript{1} score. 
This suggests that crowdworker difficulties with certain excerpts or parameters are due to more nuanced factors than length or readability.  

We  further investigate these factors by manually comparing the crowdsourced majority vote annotations to the expert annotations.
We noticed that crowdworkers had more difficulty annotating {\em senders} and {\em recipients} than {\em attributes} and  {\em transmission principles}.  Attributes and transmission principles are generally nouns or verbs, occur in lists, and require less semantic parsing to identify. In contrast, senders and attributes are often pronouns that occur singly and require more complex sentence parsing to distinguish between them.  

More detailed analysis indicated that the 160 parameter-based annotation errors fall into
four main categories. Each category has corresponding implications for crowdsourcing CI annotations.

\subsubsection{Expert Errors}
We identified 11 cases where the majority vote crowdsourced annotation was correct while the ``ground-truth'' expert annotation was incorrect. Most of these cases were due to the expert missing a one-word sender or recipient, e.g. ``we.'' We did not adjust recall or precision scores to reflect the incorrect expert annotations, as these judgments were made after, and could have been influenced by, viewing the crowdsourced annotations. 
However, the presence of these incorrect expert annotations demonstrates the non-triviality of the annotation task.

\subsubsection{Skipped Parameters} 
The most common error occurred when the crowdworkers simply neglected to annotate some or all instances of a given parameter. These errors were the primary contributor to lowering recall scores without affecting precision. We identified 117 skipped parameter errors. There are three possible reasons why crowdworkers might have neglected to annotate all instances of each parameter: 1) the workers may have considered an excerpt and honestly thought that it didn't contain the parameter, 2) the workers may have intentionally skipped entire parameters, or 3) the workers may have found one or two instances of each parameter and then moved on to the next excerpt without double-checking to ensure that none were missed. This could be due to cognitive fatigue or the fact that crowdworkers are incentivized to finish the annotations as quickly as possible to optimize their hourly compensation rate. 

As an example of reason 1, consider the sentence {\em ``We collect information when you sync non-content like your email address book, mobile device contacts, or calendar with your account.''}  Both the expert and the crowdworkers correctly labeled ``email address book,'' ``mobile device contacts,'' and  ``calendar'' as attributes. However, the expert also labeled ``information'' as an attribute, while the majority vote annotation did not. This was marked as a false negative ``skipped parameter'' error, but ``information'' does not provide any specific details about the attribute, so it is understandable that the crowdworkers omitted this label. This specific skipped parameter error (``information'' not labeled as attribute) occurred in 6 of the annotated excerpts.

Skipped errors could potentially reduced in future crowdsourcing tasks by using previous crowdworker annotations to provide ``hints'' for successive workers. For example, the number of parameters annotated by previous workers could be shown (likely as a range) to indicate approximately how many parameters the current worker should find. This would help address reason 3 for skipped errors above, providing a nudge for workers finding fewer parameters to continue searching for additional annotations. However, such hints would have to be carefully applied to prevent individual crowdworker errors from negatively influencing the collective annotation effort.

\subsubsection{Ambiguous Parameters} Ambiguous parameter errors occurred when a CI parameter was mislabeled compared to the expert annotation, but
the correct labeling is ultimately open to interpretation.
Consider the sentence {\em ``If you want to take full advantage of the sharing features we offer, we might also ask you to create a publicly visible Google Profile, which may include your name and photo.''} In this sentence, ``publicly'' could be interpreted as a recipient, i.e. the public would receive the data in the Google Profile. However, ``publicly'' could also be interpreted as a transmission principle i.e. the flow is from ``you'' to your ``Google Profile'' and the condition on the flow is that it is public. The expert labeled ``publicly'' as a recipient, while the crowdsourced majority did not.
We only identified 3 such ambiguous parameter errors, indicating that CI information flow descriptions map naturally to privacy policy texts. 

\subsubsection{Overlapping Parameters}
Overlapping parameter errors occurred when a CI parameter was mislabeled compared to the expert annotation, but the text in question is part of two or more CI parameters simultaneously.
We identified 16 overlapping parameter errors.
Consider the excerpt {\em ``When you use our services or view content provided by Google, we automatically collect and store certain information in server logs.''} 
The first clause (before the comma) could be interpreted as a single transmission principle, but the ``you'' could also be a sender. 
Variations on this issue were the primary cause of false positive errors for the ``sender'' parameter, i.e. the expert annotated an entire clause as a transmission principle but the majority vote annotation instead labeled a single word in the clause as a sender. 

The presence of overlapping parameter errors is due to a tradeoff in our implementation of the CI annotation task. 
We chose to allow only one CI parameter annotation per word in each excerpt to simplify the task for workers. 
This tradeoff could be avoided in future work by asking each crowdworker to annotate only a single CI parameter type, simplifying the task from multi-class classification to binary classification. However, this would require more crowdworkers per policy and could lead to higher rates of false positives if crowdworkers aren't forced to discriminate between different parameters. 

\subsubsection{True Errors} True errors occurred when the crowdworkers unambiguously misannotated a CI parameter. Fortunately, true errors accounted for only 13 out of 160 total errors in the majority vote annotation. This implies that when a label makes it into the majority vote annotation (with sufficient workers contributing to the vote), it is very likely to be correct. The low frequency of true errors indicates that, with improvements to reduce the number of skipped parameter errors, crowdsourcing can be a high-accuracy method of obtaining CI annotations of privacy policies.

\subsection{Summary}
\label{sec:crowdsource-discuss}

Our proof-of-concept experiment shows that crowdworkers with no prior exposure to CI are able to quickly understand and perform CI annotations of legalistic privacy policies.  
Labels which make it into a majority vote annotation compiled from several individual crowdworkers are very likely correct. 
This supports the notion that CI-style information flows are a natural way for people to think about privacy and thereby a useful framework for analyzing privacy policies and privacy policy updates. 
\section{Discussion}

We present a CI annotation methodology to help researchers and regulators assess and evaluate privacy policies. This work is a stepping stone in a larger effort to improve readability and increase transparency in disclosure of information handling practices. While philosophical in origin, the theory of CI offers a practical framework to reason about privacy implications in a given context and therefore serves as a powerful tool for reasoning about privacy preserving efforts in technical fields.

The notion of an appropriate information flow in the CI framework lends itself well to user data privacy policies; privacy statements are essentially prescribed by the policy information flows. Annotating privacy policies with CI parameter labels offers a way to apply a full-fledged formal theory of privacy to their analysis. Relevant stakeholders---consumers, legal scholars, and regulators---can perform qualitative, quantitative and normative analysis to find incomplete, vague and ambiguous privacy statements. This also enables leveraging other applications of the CI framework. For example, it is possible to compare which flows prescribed by the policy align or do not align with consumers privacy expectations~\cite{apthorpe2018discovering}. 

As privacy policies evolve, CI annotations assist comparative analyses of new updates to identify which information flows were amended, added or removed. These analyses will ideally help companies write more coherent and complete privacy policies by identifying privacy statements containing missing, vague and/or bloated CI parameters. 

Furthermore, we can use our CI annotation crowdsourcing methodology to produce a large corpus of privacy policies annotations and discover trends and patterns in the types of flows that are being prescribed by policies within and across industries. This corpus could also be used as a training set to build tools for automatically identifying CI flows and parameters in privacy policies. 

\section {Limitations and Future Work}

We have identified the following opportunities for further research to improve and streamline the CI annotation process:

First, privacy policies are not written to intentionally fit the CI framework. As discussed in Sections~\ref{sec:case_study}~and~\ref{sec:crowdsourcing},
privacy policy terms can be ambiguous, vague, compound and even missing. 
This complicates the task of annotating privacy policy text with CI parameter labels. Nevertheless, our crowdsourcing annotations showed promising results on a diverse privacy statements from privacy policies of 17 companies. In future work, we intend to continue validating the CI annotation approach on larger policy samples. 

Second, our annotation methodology deals only with statements describing information transfers. These statements comprise the majority of privacy policy text and lend themselves to the CI framework. However, other statements, such as those describing how long information is stored, when and how information is purged, and what features allow users to fine tune privacy settings, fall outside the reasoning of the CI framework. Annotating these statements will require additional methodologies to complement our approach. A blended technique, such as combining CI annotation for information transfer statements with more general tags like those used by the Usable Privacy Project \cite{sadeh2013usable}, could provide the rigor of our CI technique with the flexibility to account for the diversity of information included in privacy policies. 
\section{Conclusion}
This paper presents a methodology for analyzing privacy policies  using annotations based on the theory of contextual integrity~\cite{nissenbaum2010privacy}.  
We perform a case study annotation of pre- and post-GDPR Facebook privacy policies and demonstrate that CI offers a rigorous way to examine privacy statements. We find that Facebook's post-GDPR privacy policy describes more total information flows with more parameters than the pre-GDPR version, but the updates do not improve the percentage of flows that contain vague language, omit parameters, or include many parameters of the same type. These issues impede interpretability, preventing users from clearly understanding how their information is being collected and shared. 

To further scale our approach, we present a method for crowdsourcing CI annotation of privacy policies. We test this method on 48 excerpts from 17 policies with 141 Amazon Mechanical Turk workers. Resulting high-precision crowdsourced annotations indicate that CI annotation is an intuitive method for interpreting privacy policies and that crowdsourcing could be used to obtain a large corpus of annotated privacy policies for future analysis.

\bibliographystyle{abbrv}
\bibliography{references}

\pagebreak
\clearpage

\begin{table*}[t]
\centering
\begin{tabular}{p{2.2cm}|l|l|l}
\textbf{Statistic} & CI Parameter & \textbf{Corr. coeff.} & \textbf{p-value}\\
\hline
Total \# words & Attribute & -0.03 & 0.82 \\
& Sender & 0.03 & 0.86 \\
& Recipient & -0.11 & 0.46 \\ 
& TP & -0.15 & 0.30\\ 
&&&\\
\# words labeled & Attribute & 0.07 & 0.62 \\
as CI parameters & Sender & 0.10 & 0.48 \\
by expert & Recipient & 0.01 & 0.96 \\
& TP & -0.02 & 0.89 \\
&&&\\
Flesch-Kincaid & Attribute & 0.14 & 0.35 \\
Reading Ease & Sender & 0.20 & 0.18 \\
& Recipient & 0.10 & 0.49 \\
& TP & -0.05 & 0.76 \\
&&&\\
FOG Index & Attribute & 0.15 & 0.32 \\
& Sender & 0.19 & 0.19 \\
& Recipient & 0.10 & 0.50 \\
& TP & -0.06 & 0.67 \\
\end{tabular}
\vspace{6pt}
\caption{Spearman correlations of majority vote annotation word-based F\textsubscript{1}~scores for each CI parameter versus various statistics of corresponding privacy policy excerpts.}
\label{tbl:readability}
\end{table*}

\end{document}